# First-Principles Prediction of Novel Magnetic Materials Based on ZrCuSiAs-like Semiconducting Pnictide-Oxides

V.V. Bannikov, I.R. Shein * and A.L. Ivanovskii

*Institute of Solid State Chemistry, Ural Branch, Russian Academy of Sciences,
91, Pervomaiskaya str., Ekaterinburg, 620990 Russia*

**ABSTRACT**

We assumed that significant enlargement of the functional properties of the family of quaternary ZrCuSiAs-like pnictide-oxides, often called also as 1111 phases, which are known now first of all as parent phases for new FeAs superconductors, may be achieved by replacement of *nonmagnetic* ions by *magnetic* ions in *semiconducting* ZrCuSiAs-like phases. We checked this assumption by means of first-principles FLAPW-GGA calculations using a wide-band-gap semiconductor YZnAsO doped with Mn, Fe, and Co as an example. Our main finding is that substitution of Mn, Fe, and Co for Zn leads to drastic transformations of electronic and magnetic properties of the parent material: as distinct from the non-magnetic YZnAsO, the examined doped phases $YZn_{0.89}Mn_{0.11}AsO$, $YZn_{0.89}Fe_{0.11}AsO$, and $YZn_{0.89}Co_{0.11}AsO$ behave as a *magnetic semiconductor, a magnetic half-metal or as a magnetic gapless semi-metal*, respectively.

The discovery [1] of high-temperature superconductivity (up to $T_C \sim 56K$) in proximity to magnetism in a series of quaternary iron-based pnictide-oxides such as *LnFePnO*, where *Ln* are rare earth metals and *Pn* are pnictogens (P or As), has attracted enormous interest in this group of materials. [2-5]

Nevertheless, these Fe*Pn*-based superconductors represent a rather small group of a much broader family of layered tetragonal ZrCuSiAs-like systems, often called also as 1111 phases. At present, more than 150 such phases have been synthesized. [6-10] Besides the above pnictide-oxides, among these tetragonal 1111 phases there are also numerous pnictide-fluorides, chalcogenide-oxides, chalcogenide-flourides as well as silicide and germanide hydrides, which have a large potential for applications owing to their outstanding physical properties. [4,6-10]

Generally, all ZrCuSiAs-like phases may be divided into two main groups exhibiting a metallic-like or semiconducting behavior, respectively. The metallic-like systems (such as the above mentioned *LnFePnO*) include transition metals with partially occupied $d^{n<10}$ states, which dominate at the Fermi level and are responsible for superconductivity and magnetic effects in these materials. Oppositely, for the semiconducting systems (such as *LnZnPnO* with non-magnetic $d^{10}$ atoms) the band gap arises between bonding and antibonding states of *p* atoms.

Chemical substitutions (doping effects) are the main strategy for improvement of the properties of these systems. A large number of investigations of doping effects on *metallic* ZrCuSiAs-like phases (mainly for Fe*Pn* superconductors) have been performed to date, [2-5] in particular substitutions of *nonmagnetic $d^{10}$* ions for *magnetic $d^{n<10}$* ions. [11,12]

However, the "inverse" doping picture, namely substitution of *magnetic $d^{n<10}$* ions for *nonmagnetic $d^{10}$* ions in *semiconducting* ZrCuSiAs-like phases has not been actually examined.

In this Communication we focus on the possibility of significant enlargement of the functional properties of these advanced materials and predict the ways for designing of novel magnetic systems based on semiconductors with the ZrCuSiAs-type structure as a result of their doping with magnetic $d^{n<10}$ atoms.

As a parent system, the tetragonal arsenide-oxide YZnAsO (space group *P*4/*nmm*, Z = 2, the atomic positions are Y: 2*c*(¼,¼,$z_Y$); Zn: 2*b*(¾,¼,½); As: 2*c*(¼,¼,$z_{As}$); O: 2*a*(¾,¼,0), with optimized lattice constants: *a* = 3.9600 Å and *c* = 8.9250 Å and internal coordinates $z_Y$ = 0.1215 and $z_{As}$ =0.6816 [13]) was used, which is a wide-band-gap semiconductor with direct transition at Γ and band gap ~ 2.8 eV.[13] As is known, YZnAsO adopts a layered structure, where [YO]$^{\delta+}$ layers are sandwiched between [ZnAs]$^{\delta-}$ layers formed by edge-shared tetrahedra $ZnAs_4$.[8-10]

Then we examined the effect of substitution of some magnetic $d^{n<10}$ atoms for non-magnetic Zn atoms (inside [ZnAs] blocks). For this purpose, a 72-atomic 3×3×1 supercell of tetragonal YZnAsO ($Y_{18}Zn_{18}As_{18}O_{18}$) was chosen, and then two Zn atoms were replaced by magnetic $d^{n<10}$ atoms M = Mn($d^5$), Fe($d^6$), and Co($d^7$); in this way the nominal compositions YZn$_{0.89}$M$_{0.11}$AsO were simulated. Note that the *M* dopants were located (in the supercell) at distances $d_{M-M}$ ~ 8.4 Å, *i.e.* far apart to account for their interaction.

Our calculations were carried out by means of the full-potential method with mixed basis APW+lo (FLAPW) implemented in the WIEN2k suite of programs.[14] The generalized gradient approximation (GGA) to exchange-correlation potential in the PBE form [15] was employed. The plane-wave expansion was taken to $R_{MT}$ × $K_{MAX}$ equal to 7, and the *k* sampling with 6×6×8 *k*-points in the Brillouin zone was used. For all the examined YZn$_{0.89}$M$_{0.11}$AsO systems, in order to determine the ground state, we considered both the non-magnetic (NM) state and the magnetic state in the assumption of ferromagnetic spin configuration.

To understand the electronic properties of YZn$_{0.89}$M$_{0.11}$AsO systems, let us briefly discuss the band structure of the parent phase: quaternary arsenide-oxide YZnAsO, see also Ref. 13. Figure 1 shows the total and atomic-resolved *l*-projected DOSs for YZnAsO as calculated for equilibrium geometry. Here, the quasi-core As 4$s$ states are placed in the range from - 12 to -10 eV below the Fermi level. The fully occupied Zn 3$d$ states are located from - 8 to - 6.5 eV and are separated from the near-Fermi valence band by a gap. In turn, the valence states occupy energy intervals from the Fermi level down to - 5.2 eV, and in this region the admixtures of contributions from all atoms take place, however the near-Fermi region of YZnAsO is formed predominantly by the states of [ZnAs] blocks, see Fig. 1. The "experimental" gap for YZnAsO was estimated from our FLAPW-GGA calculations to be 2.8 eV, see also Ref. 13. The bonding picture for YZnAsO can be classified as a mixture of ionic and covalent contributions, where inside [ZnAs] and [YO] blocks, mixed ionic-covalent bonds Y-O and Zn-As take place (owing to hybridization of valence states of Y-O, Zn-As, and As-As, simultaneously with Y → O and Zn → As charge transfer), whereas between the adjacent [YO] and [ZnAs] blocks, ionic bonds emerge owing to [YO] → [ZnAs] charge transfer (about 0.3 electrons per formula unit as obtained by Bader's analysis) and weak covalent "inter-layer" Y-As bonds appear too, see Ref. 13 for details. Thus, here (similarly to the extensively investigated 1111 FeAs phases [2-5]) the oxygen-containing blocks [YO] act as "charge reservoirs".

Let us discuss the electronic properties of doped systems YZn$_{0.89}$$M$$_{0.11}$AsO. The data obtained allow us to make the following conclusions: (i) for all YZn$_{0.89}$$M$$_{0.11}$AsO, the $M$ 3$d$ bands appear in the band gap of the parent phase; and a spontaneous spin polarization (owing to the on-site exchange interaction) for all $M$ ions occurs; (ii) the splitting of $M$ 3$d$↑- $M$ 3$d$↓ bands decreases when going from Mn to Co, and (iii) a strong dependence between the type of the $M$ dopant and its atomic magnetic moment (MM) is observed, when the calculated MM are MM(Mn) = 3.57 μ$_B$, MM(Fe) = 2.82 μ$_B$, and MM(Co) = 1.85 μ$_B$. Additional magnetic moments (of ~ 0.04÷0.01 μ$_B$) are induced at the first neighbor As sites, whereas magnetization for other atoms of the parent phase is very small. Thus, the effect of magnetization of YZn$_{0.89}$$M$$_{0.11}$AsO (see also below) arises exclusively inside [ZnAs] blocks, whereas the adjacent [YO] blocks ("charge reservoirs") retain their non-magnetic character.

The most interesting result is that upon substitution of magnetic ions $M$ for zinc, the $M$ $3d$ bands are split into two spin-resolved bands, which, being partially hybridized with As states, change completely the initial picture of the electronic spectrum of the "ideal" quaternary arsenide-oxide YZnAsO, see Fig. 2. Let us comment on the effects obtained for each doped system YZn$_{0.89}$M$_{0.11}$AsO.

For YZn$_{0.89}$Mn$_{0.11}$AsO, the Mn $3d\uparrow,\downarrow$ bands adopt the maximal splitting, where Mn $3d\uparrow$ states are filled, whereas Mn $3d\downarrow$ states are free and form a new band localized inside the band gap of YZnAsO, around ~ 1 eV above E$_F$. As a result, YZn$_{0.89}$Mn$_{0.11}$AsO behaves as a *magnetic semiconductor*.

For YZn$_{0.89}$Fe$_{0.11}$AsO, the decrease in Fe $3d\uparrow,\downarrow$ spin splitting (as compared with manganese) takes place together with partial filling of Fe $3d\downarrow$ states, as depicted in Fig. 2. As a result, YZn$_{0.89}$Mn$_{0.11}$AsO is characterized by a nonzero density of carriers at the Fermi level for one spin projection (N$\downarrow$(E$_F$) > 0), but has a band gap for the reverse spin projection (N$\uparrow$(E$_F$) = 0). As is known, such kinds of spectra are typical of the so-called *magnetic half-metals* (MHMs), [16-19] for which the spin density polarization at the Fermi level is $P$ = {N$\downarrow$(E$_F$) - N$\uparrow$(E$_F$)}/{N$\downarrow$(E$_F$) + N$\uparrow$(E$_F$)} = 1. As a result, conduction in MHMs occurs along preferred spin channels, and such materials exhibiting nontrivial spin-dependent transport properties are important candidates for applications in spintronic devices.[17,18] Thus, YZn$_{0.89}$Fe$_{0.11}$AsO may be present as formed from nonmagnetic [YO] blocks and magnetic half-metallic [Zn$_{0.89}$Fe$_{0.11}$As] blocks. Let us note that a similar layered "hybrid structure" was recently observed for the related system Sr$_2$VFeAsO$_3$,[20,21] where conducting [FeAs] blocks are sandwiched between magnetic half-metallic [Sr$_2$VO$_3$] blocks.[22]

Finally, a very interesting picture arises for YZn$_{0.89}$Co$_{0.11}$AsO, where the spin-down channel adopts a *gapless semi-metal-like* spectrum, while the other spin-up channel is semiconducting-like. Thus, YZn$_{0.89}$Co$_{0.11}$AsO belongs to a rare group of very recently predicted [23] new materials: the so-called spin gapless semi-metals (or semiconductors). For these materials (as well as for "conventional" gapless semi-metals - such as graphene and some others [24,25]) no energy is required to excite electrons from the valence band to the conduction band, but unlike the above "conventional" gapless semi-metals, the excited electrons as well as holes are 100% spin polarized, and these systems will be extremely sensitive to various external effects, for example, pressure, magnetic field, electromagnetic radiation, temperature etc.

Novel physics and novel applications (first of all for spintronics), as expected for spin gapless semi-metals,[23] make the presented result for YZn$_{0.89}$Co$_{0.11}$AsO to be very attractive. Besides, Wang in Ref. 23 assumed that novel spin gapless semi-metals can be produced by doping of parent compounds - gapless semiconductors or semiconductors with a narrow direct or indirect band with magnetic ions. Our results demonstrate that the choice of initial matrixes is not limited to these gapless materials, but can be much more extensive, including a vast class of wide-band-gap semiconductors.

In summary, we have investigated the effect of doping of the non-magnetic wide-band gap semiconductor YZnAsO, which belongs to a broad family of quaternary layered ZrCuSiAs-like systems, with magnetic atoms $M$ = Mn, Fe, and Co.

Our main finding is that the replacement of Zn by Mn, Fe or Co leads to striking differences in electronic and magnetic properties of the parent phase and the doped materials, which will behave as a magnetic semiconductor (YZn$_{0.8}$Mn$_{0.11}$AsO), a magnetic half-metal (YZn$_{0.8}$Fe$_{0.11}$AsO) or as a magnetic gapless semi-metal (YZn$_{0.8}$Ni$_{0.11}$AsO). This opens up new interesting prospects for significant enlargement of the functional properties of the family of quaternary ZrCuSiAs-like systems, which can be much broader than in the presented work.

Indeed, numerous representatives of the family of semiconducting ZrCuSiAs-type compounds adopting various compositions and band gap values (such as LaZnAsO, LaCuTeO, NdCuTeO, SmCuSeF, BaZnPF, *RE*ZnPO, *RE*CuSO, *RE*AgSO, where *RE* are rare earth metals such as La, Pr, Nd, Sm, Dy etc., see Refs. 8-10) provide (*via* their doping with *various* magnetic *d* ions with *various* doping levels) a fascinating platform for further theoretical and experimental search of new magnetic materials, whose properties and potential applications can appear very exciting and interesting.

**ACKNOWLEDGMENTS**
Financial support from the RFBR (Grants 09-03-00946; 10-03-96008, and 10-03-96004) is gratefully acknowledged.

───────────────────────────

* Corresponding author.
*E-mail address:* shein@ihim.uran.ru (I.R. Shein).

# FIGURES

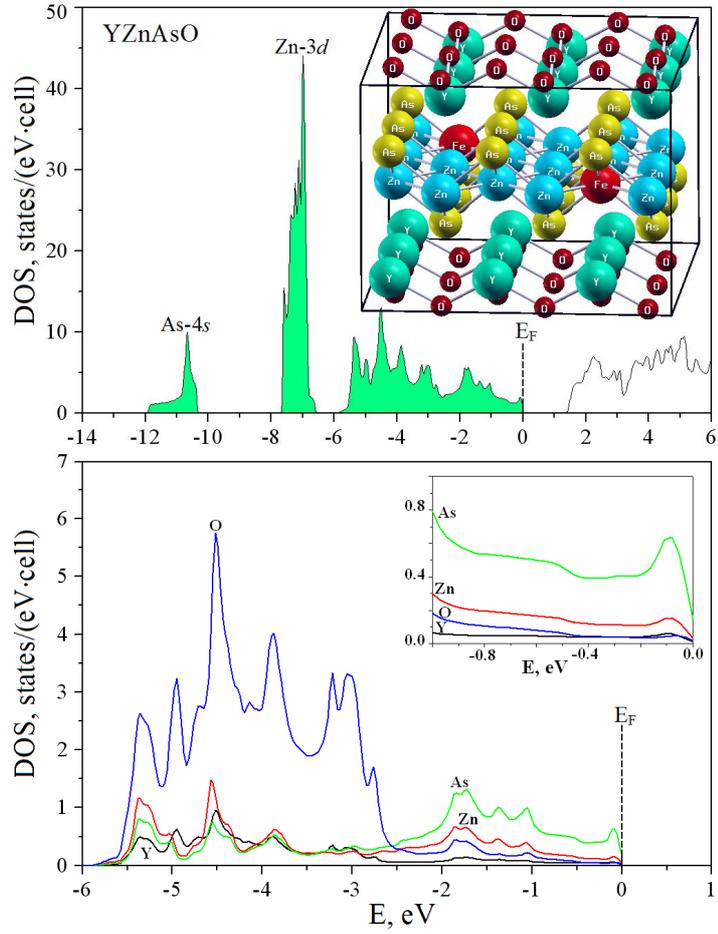

**Fig. 1.** Total (*upper panel*) and partial (*bottom panel*) densities of states for YZnAsO parent phase. *Inserts:* The crystal structure of YZnAsO supercell with introduced *M*-3*d* impurities (*upper panel*) and near-Fermi partial densities of states (*bottom panel*).

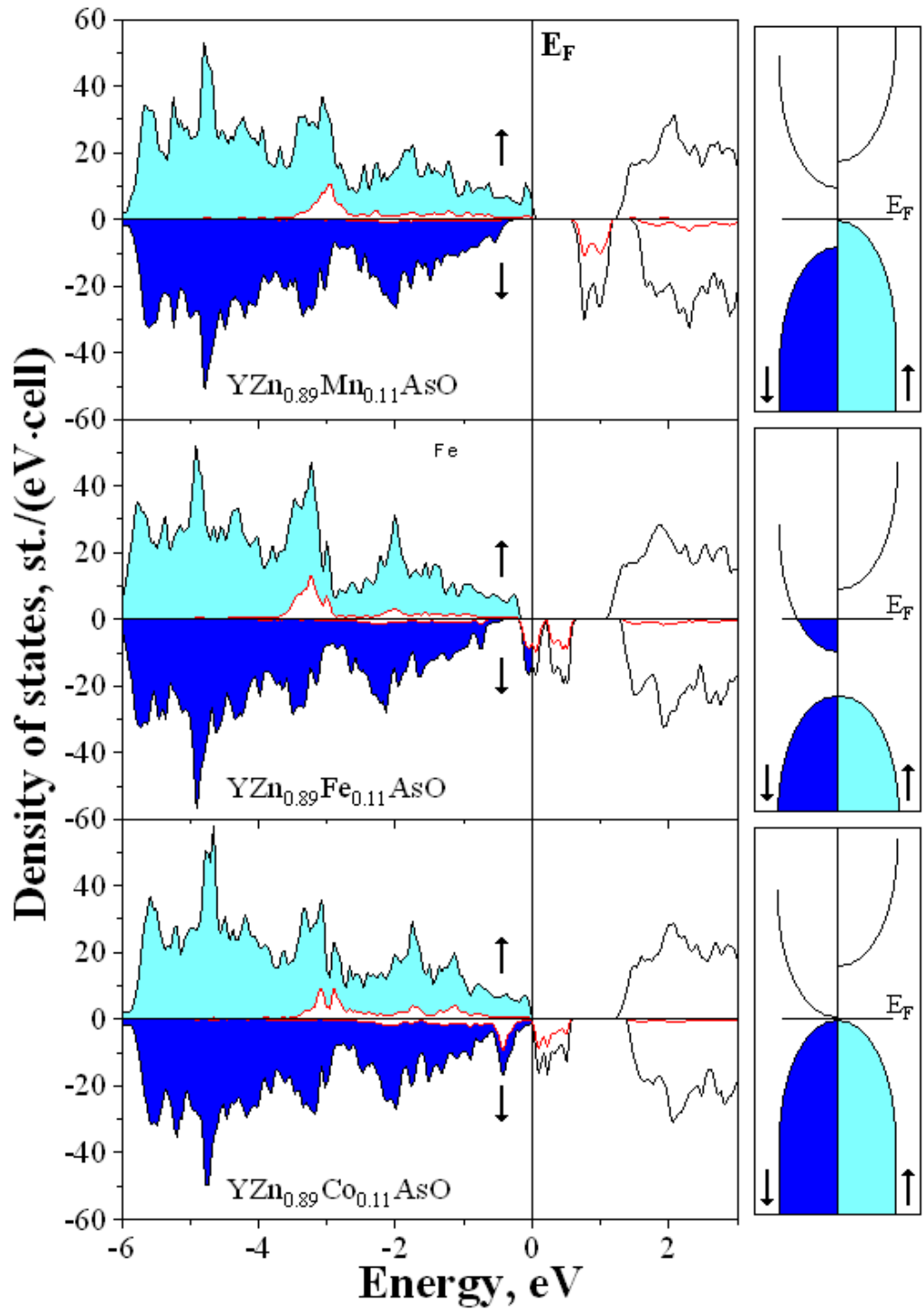

**Fig. 2**. Total spin-resolved densities of states for $YZn_{0.89}Mn_{0.11}AsO$, $YZn_{0.89}Fe_{0.11}AsO$, and $YZn_{0.89}Co_{0.11}AsO$. Partial $3d_{\uparrow,\downarrow}$ densities of states of magnetic ions are also shown as red lines. *On the right*: the schematic band diagrams for the predicted materials are depicted.